\documentclass[11pt]{article}

\usepackage{amsmath,amsmath,amssymb}

\usepackage{mathrsfs}
\usepackage{multirow}
\usepackage{graphicx}
\usepackage{authblk}
\usepackage{indentfirst}
\usepackage{multicol}  
\usepackage{tabu}
\usepackage{tabularx}
\usepackage{url}
\usepackage{fancyhdr}
\usepackage[numbers]{natbib}
\usepackage{lineno}
\usepackage[normalem]{ulem}

\usepackage{fancyhdr}
\usepackage{colortbl}
\usepackage{caption}

\usepackage{hyperref}
\hypersetup{colorlinks=true,linkcolor=blue,anchorcolor=blue,citecolor=blue}

\usepackage[table,xcdraw]{xcolor}
\usepackage{longtable}
\usepackage{hhline}
\usepackage{alltt}
\usepackage{tgcursor}
\usepackage{float}
\usepackage{xr}
\usepackage{makecell}
\usepackage{diagbox}
\usepackage{cleveref}
\makeatletter
\newcommand*{\addFileDependency}[1]{
	\typeout{(#1)}
	\@addtofilelist{#1}
	\IfFileExists{#1}{}{\typeout{No file #1.}}
}
\makeatother
\newcommand*{\myexternaldocument}[1]{
	\externaldocument{#1}
	\addFileDependency{#1.tex}
	\addFileDependency{#1.aux}
}
\myexternaldocument{supplement}

\begin{document}
	\setlength{\parindent}{2em}
	
\title{SVSBI: Sequence-based virtual screening of biomolecular interactions}
\author{Li Shen$^1$,  Hongsong Feng$^{1 }$  ,  Yuchi Qiu$^{1 }$ 
 and Guo-Wei Wei$^{1,2,3}$\footnote{
Corresponding author.		E-mail: weig@msu.edu} \\
$^1$ Department of Mathematics, \\
Michigan State University, MI 48824, USA.\\
$^2$ Department of Electrical and Computer Engineering,\\
Michigan State University, MI 48824, USA. \\
$^3$ Department of Biochemistry and Molecular Biology,\\
Michigan State University, MI 48824, USA. \\
} 
	
	\maketitle
	\begin{abstract}	
Virtual screening (VS) is an essential technique for understanding biomolecular interactions, particularly, drug design and discovery. The best-performing VS models depend vitally on three-dimensional (3D) structures, which are not available in general but can be obtained from   molecular docking. However, current docking accuracy is relatively low, rendering unreliable VS models. 
We introduce sequence-based virtual screening (SVS) as a new generation of VS models for modeling biomolecular interactions. The SVS model  utilizes advanced natural language processing (NLP) algorithms and optimizes deep $K$-embedding  strategies to encode biomolecular interactions without invoking 3D structure-based docking. We demonstrate the state-of-art performance of SVS  for four regression datasets involving protein-ligand binding, protein-protein, protein-nucleic acid binding, and ligand inhibition of protein-protein interactions and five classification datasets for the protein-protein interactions in five biological species. SVS has the potential to dramatically change the current practice in drug discovery and protein engineering.

	\end{abstract}	
	
	Key words:  Virtual screening, biomolecular interaction; natural language processing, deep learning; $K$-embedding.

 \pagenumbering{roman}

	
\pagenumbering{arabic}
	
	\section{Introduction}

Biomolecules are the building blocks of life and can be classified into various categories including carbohydrates, lipids, nucleic acids, and proteins based on their sizes, structures, physicochemical properties, and/or biological functions. Additionally, the realization of biomolecular functions is often accompanied by direct physical/chemical interactions with other biological molecules, small ligands, ions, and/or cofactors \cite{bryant2022improved}.  These interactions highly depend on the three-dimensional (3D) structures and the dynamics of molecules, as well as biomolecular conformational changes, due to their flexibility and  allostery. 
The understanding of biomolecular interactions is the holy grail of biological science.
	
 The last decade has witnessed the rapid advance in computational biology fueled by the achievement of artificial intelligence (AI) and increased computer power. With advanced techniques in data collecting, processing, analyzing, and representing, modern computational biology can study biological processes at extraordinary scales and multiple dimensions. It has achieved great success for various biological tasks   \cite{jumper2021highly,otovic2022sequential,qiu2021cluster}. The ability to understand biomolecular interactions via advanced AI approaches has a far-reaching significance to a wide range of research fields, including 
drug discovery \cite{otovic2022sequential}, 
virus prevention \cite{planas2022considerable}, 
directed evolution \cite{qiu2021cluster}, etc.   
However, the accurate and reliable prediction of biomolecular interactions is still a severe challenge. 

 Due to the inherently high correlation between structure information and molecular functions, the structure-based approaches achieved high accuracy and reliability in modeling and learning biomolecular interactions \cite{zhang2012preppi,kwon2020ak,ballester2010machine, zheng2019onionnet,cang2018representability,nguyen2020review}. As a result, current analysis and prediction of biomolecular interactions  rely heavily on the high-quality  3D  structures of interactive biomolecular complexes. Unfortunately, experimental determination of 3D structures is both time-consuming and expensive, leading to the scarcity of experimental structures, particularly, the structures of interactive biomolecular complexes.
To overcome this difficulty, molecular docking based on searching and scoring algorithms is designed to generate 3D structures of the interactive complexes, such as antibody-antigen complexes and protein-ligand complexes. Molecular docking is widely incorporated in the virtual screening (VS) of biomolecular interactions, offering an alternative means to construct the 3D structures of interactive biomolecular complexes and is a crucial step in computer-aided drug discovery (CAGD).
However, current molecular docking is prone to mistakes, rendering inaccurate 3D structures and leading to unreliable virtual screening \cite{prieto2018molecular}.
Despite the breakthrough in (non-interactive single) protein folding prediction by Alphafold2 \cite{jumper2021highly},  the structure prediction of interactive biomolecular complexes remains a severe challenge. There is a pressing need to develop innovative  strategies for the virtual screening of biomolecular interactions.  
 

 

Alternatively, sequence-based approaches may provide  efficient, robust, and easily accessible deep embeddings of biomolecular interactions without invoking 3D structure docking. 
Sequenced-based approaches are much more widely applicable than structure-based ones because the Genebank has over 240,000,000 sequences, compared to  only 200,000 3D protein structures in the Protein Data Bank (PDB), endowing sequence-based approaches much boarder applicability. There are  three major types of sequence-based approaches: (1) composition-based methods such as amino acid composition (AAC) \cite{zhou2008differences}, nucleic acid composition (NAC) \cite{zhao2022protein}, and pseudo AAC (PseAAC) \cite{chou2009pseudo}; (2) autocorrelation-based methods such as auto-covariance \cite{zeng2009using}; and (3) evolution-based methods such as position-specific frequency matrix (PSFM) and position-specific score matrices (PSSM) \cite{chou2009pseudo}.
Composition-based methods construct embeddings based on the distribution of single residues or substrings. Autocorrelation-based methods are based on statistical measurement of physicochemical properties of each residue, such as hydrophobicity, hydrophilicity, side-chain mass, polarity, solvent-accessible surface area, etc. Evolution-based methods extract the evolutionary information from large databases by evaluating the occurrence of each residue or the score of that residue being mutated to another type. These methods usually outperform composition-based and autocorrelation-based methods due to their efficient use of a large number of molecular sequences selected by billions of years of natural evolution. Natural language processing (NLP) based methods have been widely used to embed molecules. Among them,  autoencoders (AE),  long short-term memory (LSTM), and  Transformer are most popular.  A LSTM model, UniRep, provides enables sequence-based  rational protein engineering  \cite{alley2019unified}. An in-house autoencoder was trained with 104 million sequences \cite{feng2022machine}. 
Evolutionary scale modeling (ESM) is a large-scale Transformer trained on 250 million  protein sequences, which achieved state-of-art performance in many tasks, including structure predictions \cite{rives2021biological}. For DNA  in the genome,   pre-trained bidirectional encoder representation  model DNABERT has achieved  success in non-coding DNA tasks, such as the prediction of promoters, splices, and transcription factor binding sites \cite{ji2021dnabert}. Furthermore, an in-house small molecular Transformer was trained with over 700 million sequence data \cite{chen2021extracting}. However, none of these methods was designed for biomolecular interactions.


In this work, we proposed a novel sequence-based visual screening (SVS) of biomolecular interactions that can predict a wide variety of biological interactions at structure-level accuracy without invoking 3D structures. The biological language processing module in SVS consists of multiple NLP models, extracts evolutionary, and contextual information from different biomolecules simultaneously to reconstruct sequence representations for interactive molecules, such as proteins, nucleic acids, and/or small molecules. SVS has a strong generalizability to various types of tasks for biomolecular properties and interactions. In particular, SVS  provides the optimal $K$-embedding  strategy to study the interactions between multiple (bio)molecules with negligible computational cost. The intramolecular patterns and intermolecular mechanisms can be efficiently captured by our  SVS  without performing the expensive and time consuming 3D structure-based docking. We showed the cutting-edge performance of SVS on nine prediction tasks, including  four  binding affinity scoring functions (i.e., protein-ligand, protein-protein, protein-nucleic acid,  and ligand inhibition of protein-protein interactions) and five classification datasets for protein-protein interactions (PPIs). 
Extensive validations indicate that SVS is a general, accurate, robust, and efficient new method for the virtual screening of biomolecular interactions.

\section{Results}
\subsection{Overview of the SVS  framework}

\begin{figure}[htb!]
	\centering
	\includegraphics[width=6in]{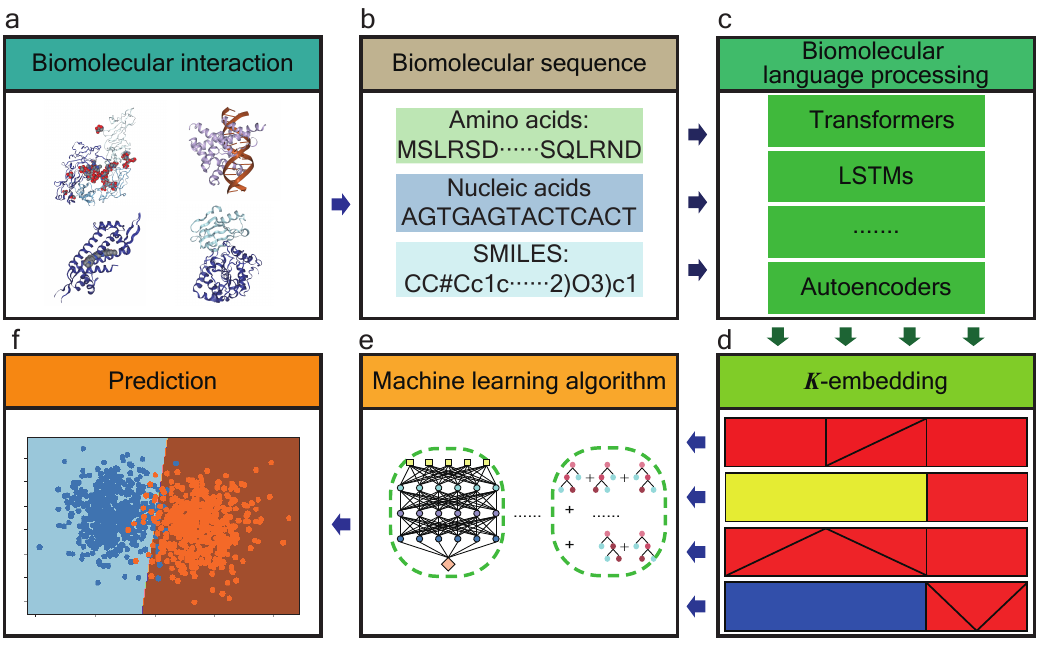} 
	\caption{Methodological workflow of SVS. 
(a) SVS is designed for a wide variety of biomolecular interactions involving proteins, DNA, RNA,  ligands, and their arbitrary combinations.  
(b) Molecular sequences are extracted from proteins, nucleic acids, and small molecular ligands involved in biomolecular interaction complexes. 
(c) The biomolecular language processing module presents the NLP embeddings of  biomolecular complexes from sequence information. 
(d) The $K$-embedding  module generates the optimal $K$-embedding of  biomolecular interactions from  the lower-order NLP embeddings. 
(e) Supervised machine learning algorithms learn from the optimal $K$-embedding model of biomolecular interactions. In principle, there are no restrictions on the choice of algorithms. Specifically, in this work, we use GBDT and ANN. 
(f) Machine learning algorithms are applied to various classification and regression tasks, including membrane protein classifications, therapeutic peptide identifications, protein-protein interaction identifications, binding affinity prediction of protein-protein, protein-ligand, protein-nucleic acids interactions, and inhibition of protein-protein interaction.}\label{fig:fig_con}
\end{figure}
 
Our SVS is a sequence-based framework offering deep learning predictions of biomolecular interactions (\autoref{fig:fig_con}). 
First, the biomolecular interaction module identifies { types of interactive biomolecular partners and treats the problem in the corresponding flow. }
Then,  the related sequences are collected and curated in the biomolecular sequence module.  
Additionally, the biomolecular language processing module generates the NLP embeddings of individual interactive molecules from  their sequence data. 
Moreover, the $K$-embedding  module further  engineers interactive $K$-embeddings from individual NLP embeddings to infer their interactive information. 
Last, the downstream machine learning algorithm module offers the state-of-the-art regression and classification  predictions of various biomolecular interactions.


In the biological language processing module, NLP embeddings are generated   for proteins, nucleic acids, and small molecules using their sequence data (\autoref{fig:fig_con}b). We employ various types of NLP models including protein LSTM model (UniRep) \cite{alley2019unified}, protein  Transformer (ESM) \cite{rives2021biological}, DNA  Transformer (DNABERT) \cite{ji2021dnabert}, small molecular Transformer \cite{chen2021extracting}, and small molecular autoencoder \cite{feng2022machine}. We particularly focus on Transformer models due to their state-of-art performance with the consideration of sequence dependencies via an attention mechanism \cite{vaswani2017attention, devlin2018bert,chen2021algebraic}. Enrich information, such as evolutionary information, 3D structure, and biochemical properties \cite{rives2021biological,chen2021extracting} can be inferred by Transformers. 

 The $K$-embedding  module (\nameref{sec_method} \ref{sec:k_emb}) takes multiple embeddings from interactive molecular components as inputs and integrates them into an optimal deep $K$-embedding model to decipher biomolecular properties and intermolecular interactions   (\autoref{fig:fig_con}d). The traditional 3D structure-based virtual screening models require a molecular docking procedure to generate the 3D molecular structures of the interactive complexes, which is inefficient and unreliable  \cite{ramirez2016reliable}. The accuracy and effectiveness of a structure-based docking method are jointly determined by multiple sub-processes including molecular structure determination \cite{bryant2022improved}, rigid and flexible docking space search \cite{bryant2022improved}, and scoring function construction \cite{jain2006scoring}. Current studies have achieved success in each of these sub-processes. However, minor errors in these sub-processes may accumulate and result in unreliable structure-based docking. Alternatively, in our SVS framework, the $K$-embedding  strategies can convert the distribution information of interactive molecular embeddings into the optimal $K$-embedding  and extract essential characteristics of biomolecular interactions, which enhances the modelability of machine learning algorithms in learning hidden nonlinear molecular interactive information. 
 
The machine learning module takes the   $K$-embedding strategies from the $K$-embedding  module for molecular property predictions. The downstream machine learning algorithms include artificial neural network (ANN) and  gradient boost decision tree (GBDT)   for predictive tasks. The hyperparameters of both models are systematically optimized via Bayesian optimization or grid search to accommodate for different sizes of datasets and deep $K$-embeddings, and different tasks (\nameref{sec_method} \ref{sec:models} and \ref{sec:bayes_opt}).

\subsection{Biomolecular binding affinity predictions}\label{Sec:BindingAffinity}

Quantitatively, binding affinity, defined as the strength of molecular interactions, is reflected in the physicochemical terms of dissociation constant ($\text{K}_d$), inhibitor constants ($\text{K}_i$), half maximal inhibitory concentration ($\text{IC}_{50}$), or corresponding Gibbs free energy \cite{steinbrecher2010towards}. Accurate predictions of molecular binding affinities are not only an important step in modeling biological systems but also a fundamental issue  for several practical usages including drug discovery \cite{ballester2010machine,cang2018representability,meng2021persistent}, 
molecular engineering,  
and mutagenesis analysis \cite{qiu2021cluster}. 

\subsubsection{Protein-ligand binding scoring} \label{Sec:Protein-ligand}

 The scoring of protein-ligand binding complexes is the ultimate goal of virtual screening in drug discovery. Typically, millions of drug candidates are screened for a given drug target. The accuracy and efficiency of virtual screening are essential for drug discovery \cite{ballester2010machine,pan2022aa}. Currently, inaccurate 3D structure-based docking and the associated unreliable virtual screening are the main obstacles in rational drug design and discovery. 
\begin{figure}[htb!]
	\centering
	\includegraphics[width=6in]{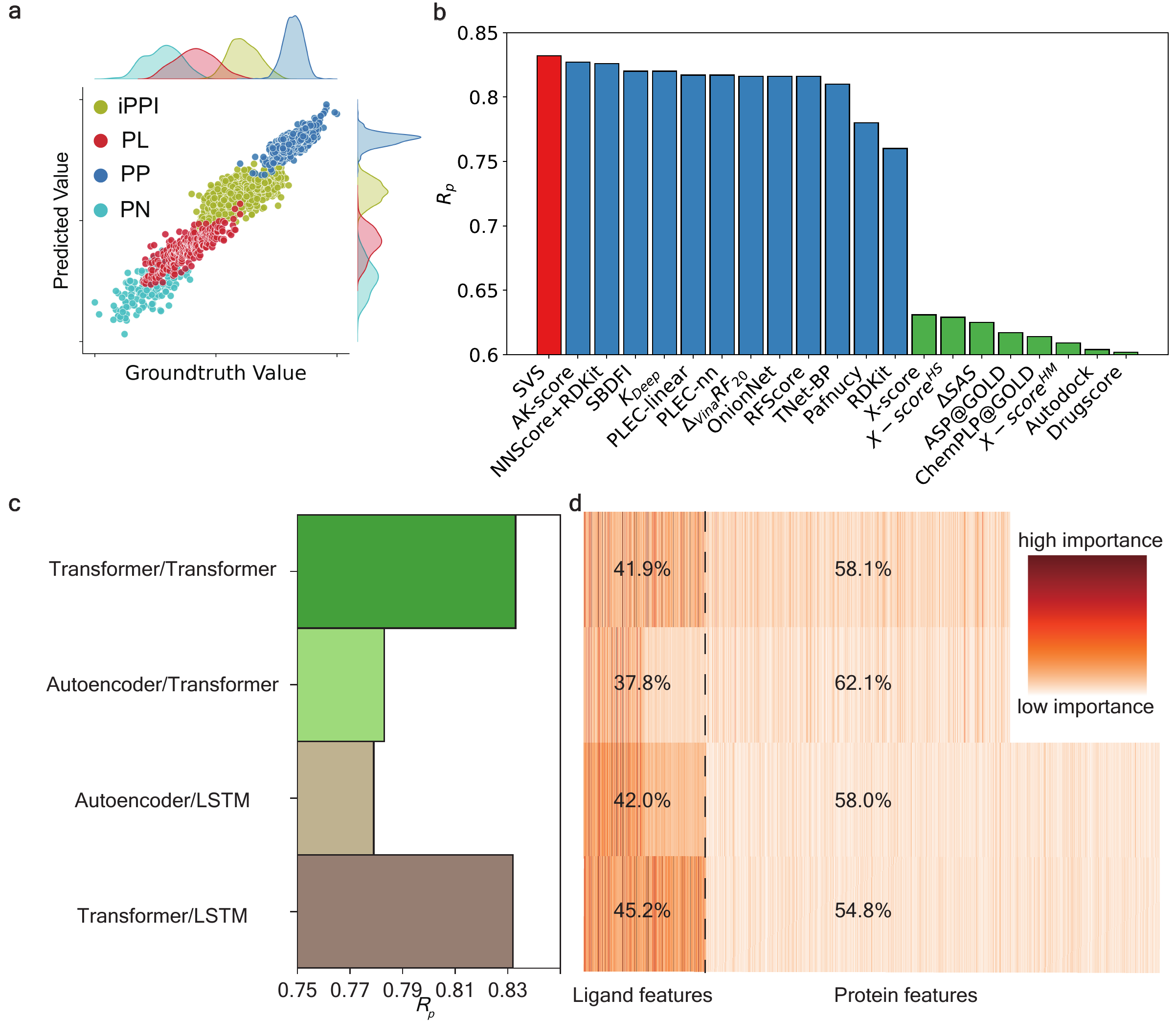} 
	\caption{ (a) A comparison of scaled predicted binding affinities and experimental results for the binding affinity predictions of protein-ligand (PL), protein-nucleic acid (PN), protein-protein (PP),  and the inhibition of PPI (iPPI) datasets. Each dataset is scaled to a specific region with an equal range for clear visualization. 
(b) Comparison of the Pearson correlation coefficient ($R_\text{p}$) of our SVS model and that of other structure-based approaches for the protein-ligand binding affinity prediction of the PDBbind-2016 core set \cite{su2018comparative}. 
Results in red, blue, and green colors are obtained using no structure (i.e., sequence), experimental structures, and docking generated structures of protein-ligand complexes, respectively.  
 Our SVS outperforms the state-of-the-art models, such as AK-score\cite{kwon2020ak}, NNScore+RDKit \cite{boyles2020learning}, and many others \cite{ jimenez2018k,wojcikowski2019development,zheng2019onionnet,stepniewska2018development,su2018comparative,jones2021improved}.
(c) Comparison of different NLP models for the Pearson correlation coefficients $R_\text{p}$ of the protein-ligand binding prediction. 
(d) The relative importance distributions of different NLP models as shown in (c). Each row consists of 512+1280/1900 colored vertical line, and each represents the importance of one feature that is generated by the  NLP models. The black dashed line is the dividing line for features belonging to different type of  molecules. The percentage on the left or the right of the black dashed line is the proportion of the summation of importance of features for the same type of molecules.}	\label{fig:pl}
\end{figure}

In this study, we applied SVS to predict the protein-ligand binding affinity on the PDBbind 2016 dataset \cite{su2018comparative}, a popular benchmark dataset employed by hundreds of research teams to validate their protein-ligand binding scoring functions \cite{kwon2020ak,jimenez2018k,wojcikowski2019development,ballester2010machine,zheng2019onionnet,stepniewska2018development,su2018comparative,su2018comparative,jones2021improved,boyles2020learning}. It  has the training data of 3772 protein-ligand complexes from the PDBbind 2016 refined set  and the test data of 285 complexes from the core set.  The availability of 3D complex structures in PDBbind database favors structure-based scoring functions, such as algebraic topology-based machine learning models, such as TopBP \cite{cang2018representability}, PerSpect-ML\cite{meng2021persistent}, and AA-score \cite{pan2022aa}.  
 
 The best performance of 2D fingerprint-based methods, achieved by the protein–ligand extended connectivity (PLEC) fingerprint \cite{wojcikowski2019development}, was $R_\text{p}=0.817$. In fact, 3D structure information was utilized in PLEC, highlighting the importance of 3D structures in existing protein-ligand binding scoring functions.  We select this dataset to examine whether the proposed SVS, without resorting to structural information,  can reach the same level of accuracy as structure-based scoring functions.    
 
As shown in \autoref{fig:pl}b, our SVS model gives the accurate prediction of binding affinity with $R_\text{p}=0.832$ and RMSE 1.696 kcal mol$^{-1}$ (\autoref{fig:pl}b). For structure-based methods, $R_\text{p}>0.7$ can be usually achieved  if experimental structures of protein-ligand complexes are used, while lower $R_\text{p}< 0.65$ is achieved when molecular docking, such as ASP$@$ GOLD and Autodock, is used to generate the structures of protein-ligand complexes  \cite{su2018comparative}. 
The structure-based TopBP method, using algebraic topology to simplify the structure complexity of 3D protein-ligand complexes, achieved the best performance with $R_\text{p}$/RMSE of  0.861/1.65 kcal mol$^{-1}$ \cite{cang2018representability}. Excluding advanced mathematics-driven structure-based methods, SVS  outperforms other structure-based methods, e.g., AK-score\cite{kwon2020ak} ($R_\text{p}$: 0.827),
NNScore+RDKit \cite{boyles2020learning} ($R_\text{p}$: 0.826)
 (\autoref{fig:pl}b). This achievement is of enormous significance that the quality and reliability of the current virtual screening can be dramatically improved to the level of x-ray crystal structure-based approaches without depending on 3D experimental structures. Our result has a far-reaching implication --- reliable virtual screening can be carried out on any drug target without relying on the 3D structures of drug-protein complexes.

  The performance from different combinations of protein and ligand embeddings are further explored (\autoref{fig:pl}c). 
	We used ESM Transformer \cite{rives2021biological} and UniRep LSTM \cite{alley2019unified} model for protein embedding, and a Transformer \cite{chen2021extracting} and an autoencoder \cite{feng2022machine} model for ligand embedding. Our analysis indicates that the small molecular Transformer outperforms the autoencoder. Additionally, Transformer achieves better performance than LSTM model for protein embedding. Further feature analysis is provided from the feature importance analysis from GBDT (\autoref{fig:pl}d). Both small molecular embeddings have the dimension of 512. For the protein embeddings, Transformer dimension is 1280, and LSTM is 1900. First, small molecular features have more highly important ones. The average importance of small molecular features are 0.082 (41.9/512), 0.074, 0.082, and 0.088 for four cases from top to bottom (\autoref{fig:pl}d). In contrast, the average importance of protein features are 0.045, 0.049, 0.031, and 0.028 for four cases.   
Additionally, the small molecular Transformer offers more important features than the autoencoder does. For the protein embeddings, the Transformer has more important features than the LSTM does. Therefore, the combination of the ligand Transformer  and  protein ESM Transformer achieves the best prediction as shown in  \autoref{fig:pl}c.

\subsubsection{Protein-protein binding affinity prediction}

Protein-protein binding affinity refers to the strength of the attractive interaction between two proteins, such as an antibody-antigen complex, when they bind to each other. It is  important metric for assessing the stability and specificity of protein-protein interactions (PPIs), which are vital for many biological processes.

Understanding protein-protein binding affinity is important for many applications, including drug discovery, antibody design, protein engineering, and molecular biology. For example, knowing how antibody-antigen binding affinity is affected by the shape of the antibody, the charge and hydration of the antibody, and the presence of specific binding sites or residues on the antibody, one can engineer antibodies with specific binding properties to neutralize viruses \cite{wang2020topology,liu2022hom}.

The protein-protein binding affinity can be quantified by Gibbs free energies. The surface plasmon resonance (SPR), isothermal titration calorimetry (ITC), 
enzyme-linked immunosorbent assay (ELISA), and Western blotting  are used to determine protein-protein binding affinities. In our work, we build a SVS   model to predict protein-protein binding affinities from protein sequences. 
We collect and curate a set of  1795 PPI complexes (\nameref{sec_method} \ref{sec:dataset}) in the PDBbind database \cite{liu2015pdb}. This dataset is employed to show the versatile nature of SVS.  Sequences of these PPI complexes are extracted and embedded using the transformer. The PPIs are represented by the stack of their Transformer  embeddings in our study. Our SVS model reached the $R_\text{p}$ of 0.743 and the RMSE of 1.219 kcal mol$^{-1}$ via 10-fold cross-validation, and the comparison of predicted value versus the ground truth  is shown in \autoref{fig:pl}a.
Our result indicates SVS is a robust approach for predicting the binding affinity of PPIs.  

\subsubsection{Protein-nucleic acid binding affinity prediction}

Another class of biomolecular interactions is protein-nucleic acid binding which plays important roles in the structure and function of cells, including catalyzing chemical reactions, transporting molecules, signal transduction, transcription, and translation.   It is also involved in the regulation of gene expression and in the maintenance of chromosome structure and function. Dysregulation of protein-nucleic acid binding can lead to various diseases and disorders, such as cancer, genetic disorders, and autoimmune diseases.   The understanding of the factors, such as hydrogen bonding, dipole, electrostatics, 
Van der Waals interaction, hydrophobicity, etc. that influence protein-nucleic acid binding affinities can be utilized to design new therapeutic molecules.  
 
In this work,  we apply SVS to analyze and predict protein-nucleic acid binding affinity. Due to the lack of existing benchmark datasets, we extract a dataset from the PDBbind database \cite{liu2015pdb}.  A total of 186 protein-nucleic acid complexes was collected (\nameref{sec_method} \ref{sec:dataset}). This dataset is chosen to demonstrate that the SVS works well for predicting nucleic acid-involved biomolecular interactions.  For this problem, our  SVS utilizes a Transformer (ESM-1b) for embedding protein sequences and another Transformer (DNABERT) for embedding nucleic acid sequences.   Our model shows good performance with an average $R_\text{p}$/RMSE of 0.669/1.45 kcal mol$^{-1}$ in a 10-fold cross-validation. Our results are depicted in  \autoref{fig:pl}a. 
Considering the fact that the dataset is very small, our SVS prediction is very good. 

\subsubsection{Inhibition of protein-protein interaction prediction}

Having demonstrated SVS for protein-ligand, protein-protein, protein-nucleic acid binding predictions, we further consider a problem involving multiple  molecular components. The small molecule inhibition of protein-protein interaction prediction (iPPI) involves at least three molecules. 

Protein-protein interactions are essential for living organisms. Dysfunction of PPIs can lead to various diseases, including immunodeficiency, autoimmune disorder,  allergy, drug addiction,  and cancer \cite{rodrigues2021pdcsm}.  Therefore, the inhibition of PPIs (iPPIs) is of great interest in drug design and discovery.  Recent studies have demonstrated significant biomedical potential for iPPIs with ligands \cite{jubb2015flexibility}.   

However,  iPPI  with ligands is challenging in a vast range of investigation phases including target validation, ligand screening, and lead optimization \cite{laraia2015overcoming}. Traditional computational methods for iPPI predictions have various limitations. For example,  structure-based approaches have to overcome the complexity of ligand docking caused by the large and dynamic interfaces of PPIs even with stable and reliable experimental complex structures \cite{watkins2015structure}. Recently, Rodrigues et al. \cite{rodrigues2021pdcsm} have developed an interaction-specific model, called pdCSM-PPI, which utilizes graph-based representations of ligand structures in the framework of ligand-based virtual screening. An important characteristic of their approach is that their models are ligand-based and target-specific: the input of each model is a set of ligands that target one particular PPI. Instead of exploring the hidden mechanism of iPPI, their models rely on a comparison of ligands by assuming that ligands with similar structures exhibit similar behavior, i.e., the similar property principle. Their approach avoids the difficulties of lacking iPPI structures and molecular mechanisms by using target-specific predictions, in which one machine-learning model is built for ligands targeting the same PPI system. Therefore, it cannot be used for the screening of new targets.  By contrast, SVS can avoid this difficulty by sequence embedding of PPI targets.  As a result,  SVS can be directly applied to explore the inhibition of new PPIs without matching targets in existing iPPI datasets. 

\begin{figure}[htb!]
	\centering
	\includegraphics[width=6.2in]{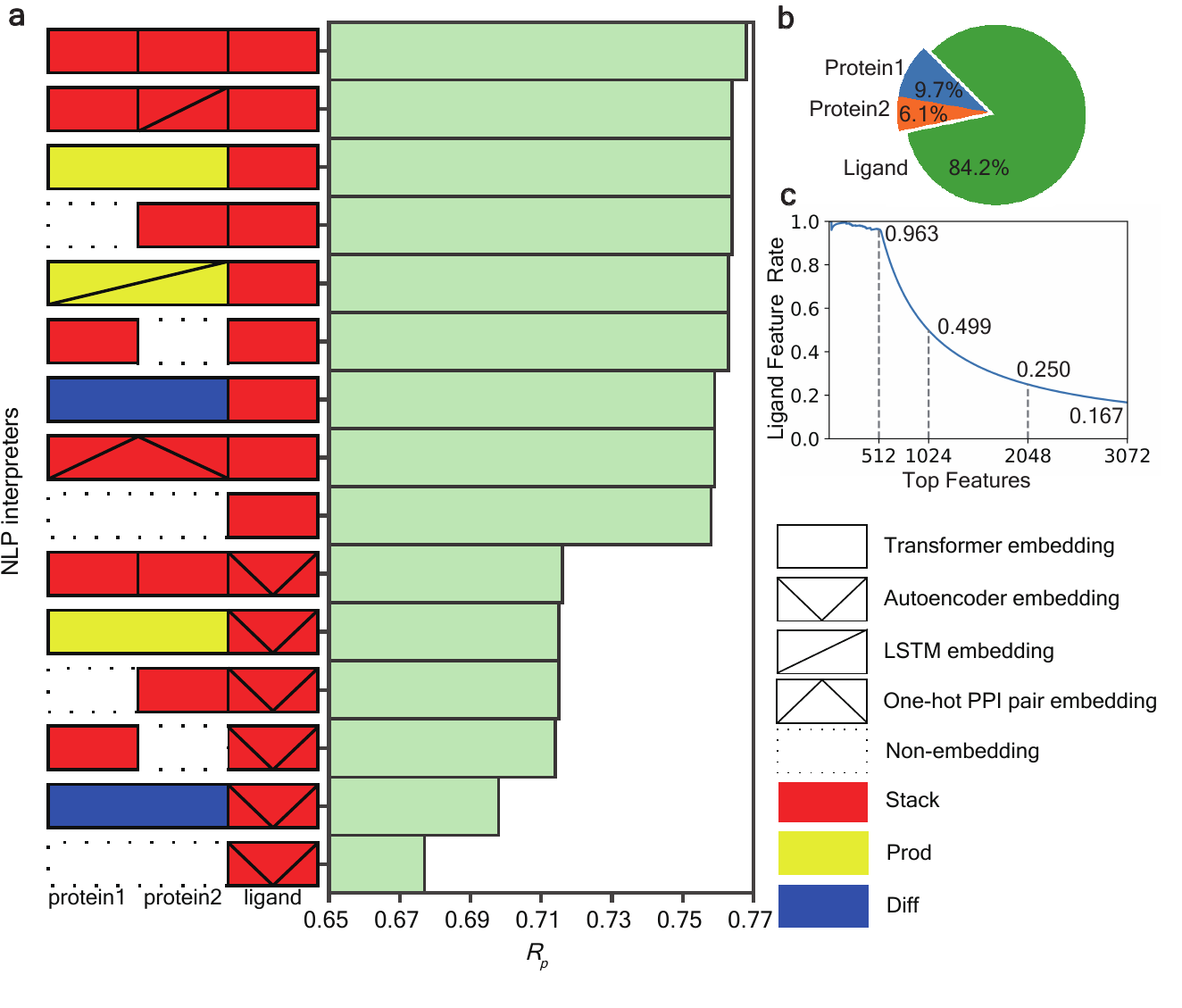} 
	\caption{(a) Illustration of the performances  ($R_\text{p}$) of various  $K$-embedding  strategies. 
	(b) The feature importance analysis of ligand, protein1, and protein2 in iPPI predictions using the best  $K$-embedding  strategy (i.e., the stack of three Transformers). 
	(c) The proportion of ligand features in top features of SVS for iPPI using the best  $K$-embedding  strategy (i.e., the stack of three Transformers). The $x$-axis indicates the quantity of top features to be considered and the $y$-axis represents the proportion of ligand features in the top features.}	
	\label{fig:ippi}
\end{figure}

In this work, we analyzed PPIs and ligands by using various $K$-embedding  strategies, to predict the half-maximal inhibitor concentration (IC50) of the ligand inhibition of PPI. For each iPPI complex, a small molecular Transformer and a protein Transformer are used to embed one ligand sequence and two protein sequences in our  SVS. 
We tested our model on the dataset considered by Rodrigues et al. \cite{rodrigues2021pdcsm}. Our model shows an $R_\text{p}$ of 0.766 and RMSE of 0.761 mol/L in the 10-fold cross-validation, while the $R_\text{p}$ and RMSE of the earlier pdCSM-PPI model are 0.74 and 0.95 mol/L, respectively. SVS shows a better performance in both $R_\text{p}$ and RMSE, illustrating the superiority of the SVS method. The comparison of predictive results versus the ground-truth value of our model can be found in \autoref{fig:pl}a.

We explore $K$-embedding strategies via various NLP deep embeddings. We examine three integrating functions in this study, i.e., \textit{Stack}, \textit{Prod}, and \textit{Diff}, to generate $K$-embedding strategies  with the higher-order embedding built from lower-order embeddings. 
 For simplicity, Stack is to concatenate two biomolecular language processing embeddings from two proteins in a PPI complex into a  single embedding vector. This method preserves the complete information provided by the biomolecular language processing module, but the downside is its high dimensionality. Since two proteins in a PPI complex are encoded by two vectors of identical length, $2$-embedding  can be done via the component-wise operations between these two vectors. We also tested the component-wise product (\textit{Prod}) and the absolute value of the difference (\textit{Diff}). These   component-wise 2-embedding approaches result in lower-dimensional 2-embeddings for the downstream machine-learning module. The specific formulas corresponding to these three strategies are described in \autoref{eq:stack}, \autoref{eq:prod}, and \autoref{eq:diff}, respectively.

Here, we choose 14 kinds of higher-order deep embeddings that take the full consideration of the homogeneity or heterogeneity of NLP models, which are shown in \autoref{fig:ippi}a with their predictive performance. It is worth noting that this iPPI dataset is a ligand-central dataset consisting of multiple ligands that target the same PPI. Therefore, $1$-embedding  for ligand sequence information processing will play the most important role. Our experiments show that using Transformer-based models with the \textit{Stack} schemes will give a state-of-the-art performance.

We further analyze the feature importance of our best schemes from GBDT for features encoding ligands and proteins. Interestingly, features for ligands are significantly more important than that for proteins (\autoref{fig:ippi}b). Specifically, the importance for ligand features is much higher at 84.2\%, while the sum of importance for two proteins is only 15.8\%. On the other hand, top features include a high proportion of ligand features, for example, 96.4\% of the top 512 features are from ligand features (\autoref{fig:ippi}c). A possible reason for such feature imbalance may be because only a few PPI systems are included in this dataset which has 1694 ligands but only 31 PPIs. Despite protein features being less important, they are necessary for learning iPPI without matching targets. As shown in \autoref{fig:ippi}a, without PPI information (non-encoding of PPIs), or with only trivial classification information of PPI (one-hot pair encoding of PPIs), our models show a significant decline in the predictive accuracy. The only exception is \textit{Diff} of the PPI target. One reason is that many proteins in this PPI target belong to the same protein family. Thus, the high similarity of these proteins in sequence would only provide very limited information for \textit{Diff} schemes. In general, the protein features are necessary components for learning target-unmatched iPPIs.

 \subsection{Protein-protein interaction identification}
 
 Protein-protein interactions (PPIs) regulate many biological processes, including signal transduction, immune response, and cellular organization \cite{sun2017sequence}. However, the selectivity and strength of PPIs depend on species and the cellular environment.  Identifying and studying PPIs can help researchers understand the molecular mechanism of protein functions and how proteins interact with one another within a cell or organism.

 \begin{figure}[htb!]
 	\centering
 	\includegraphics[width=5.5in]{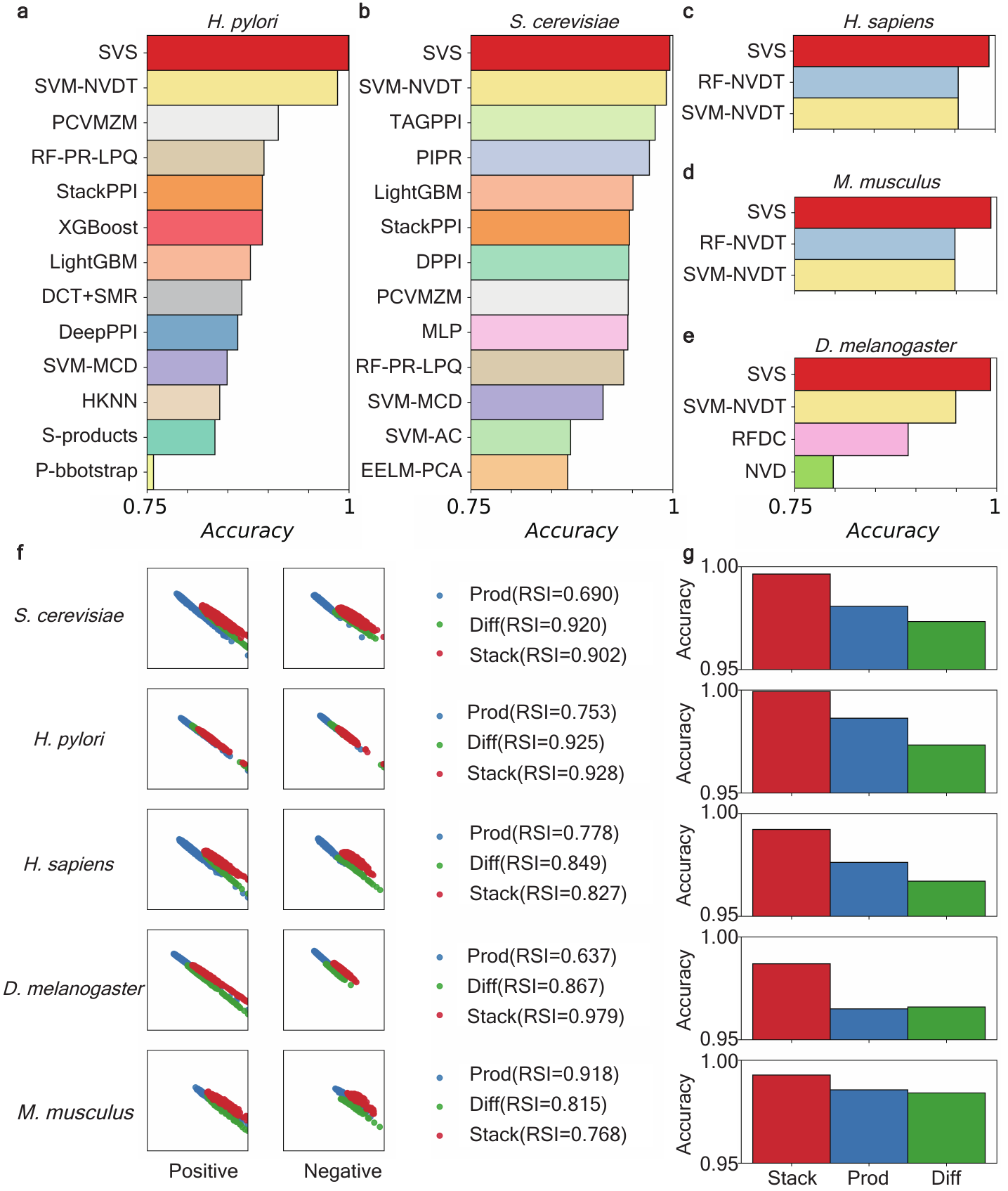} 
 	\caption{Performance analysis of the SVS for five Protein-protein interaction and non-interaction classification  datasets. (a)-(e) Comparisons of our predictive model (SVS) with some previous PPI identification models. The comparison of each dataset is shown independently in a subplot with the name of the dataset at top of it. For each subplot, the $x$-axis represents the accuracy scores, ranging from 0.75 to 1; the $y$-axis lists the name of each model. Our SVS outperforms the state-of-art models, such as SVM-NVDT \cite{zhao2022protein}, RF-NVDT\cite{zhao2022protein}, PCVMZM\cite{wang2017pcvmzm}, TAGPPI\cite{song2022learning}, etc. 
(f) Comparison of different $K$-embedding  strategies, measured by R-S analysis on features. Three $K$-embedding  strategies, \textit{Prod}, \textit{Diff}, and \textit{Stack}, are chosen for comparison. This plot is vertically composed of five similar sections. Each section represents a dataset with the name on the left. Furthermore, each section possesses two parts. The left part has two subplots showing the R-S plot of positive or negative features generated by different strategies. The right part shows the R-S Index (RSI) of different strategies. 
(g) The comparison accuracy   of predictive models of different $K$-embedding  strategies. }	\label{fig:ppi}
 \end{figure}

 We applied the SVS method to determine whether a pair of proteins interact or not.  
 Five PPI datasets with different species including \textit{Homo sapiens} (HS), \textit{Mus musculus} (MM), \textit{Saccharomyces cerevisiae} (SC), \textit{Drosophila melanogaster} (DM), and \textit{Helicobacter pylori} (HP) are employed for the benchmark.  
 Here, we explore three $K$-embedding  strategies: \textit{Stack}, \textit{Prod}, and \textit{Diff}.  

 Since the performance of regression models is complicated, we first analyze the performance of interactive features without downstream regression models. In particular, we employed the R-S plot to visualize feature residue score (R) versus similarity score (S) \cite{hozumi2022ccp}. 
 The R-score and S-score of a given sample are calculated by considering the distances of its features with that of inter-class samples and intra-class samples, formulated as \autoref{eq:r-score} and \autoref{eq:s-score}, respectively. Both R-score and S-score range from 0 to 1. A sample with a higher R-score indicates that it is far from samples in other classes, and a higher S-score indicates that it is close to other samples in the same class. An effective featurization method is expected to have both high R-scores and S-scores, despite a clear trade-off exists between R- and S-scores (\autoref{fig:ppi}b). Notably, such a trade-off can also be quantified by the R-S index (\autoref{eq:rsi}). The R-S analysis shows that \textit{Stack} features are located at the upper right of \textit{Prod} and \textit{Diff} embeddings except for the \textit{H. pylori} dataset (located in a similar area), though they overlap extensively over all datasets. In addition, from the perspective of the R-S index, \textit{Stack} and \textit{Diff} have advantages in two datasets, and \textit{Prod} has advantages in one dataset. 
 
 Furthermore, we compared different $K$-embedding  strategies by coupling with the identical regression models using five-fold cross-validation (\autoref{fig:ppi}b). 
 Consistently, the \textit{Stack}  strategy showed the highest accuracy score than others in their downstream model performance for all datasets tested (\autoref{fig:ppi}c). Overall, \textit{Stack}   provides an optimal $K$-embedding  strategy. 
 
 Overall, our models with the best \textit{Stack} of biomolecular language processing embeddings showed accuracy scores as high as 99.93\%, 99.28\%, 99.64\%, 99.22\%, and 98.69\% for datasets 
\textit{Helicobacter pylori}, 
\textit{Mus musculus}, 
\textit{Saccharomyces cerevisiae}, 
\textit{Helicobacter pylori}, and 
\textit{Drosophila melanogaster}, 
respectively (\autoref{fig:ppi}a and Table S1).  
In comparison, the state-of-art method, SVM-NVDT \cite{zhao2022protein}, gives 
98.56\%, 
94.83\%   
99.20\%,  
95.41\%,  and 
94.94\%,   respectively
for these datasets.  SVM-NVDT  was based on natural vectors and  dinucleotide and triplet nucleotide information.   
 Also, we display the AUC curves (1.00 for all cases) of our models in Figure S1. Our models outperform all previous models by a significant margin, which demonstrates the superiority of our method over previous methods for identifying PPIs.

\section{Discussions}


\begin{figure}[htb!]
	\centering
	\includegraphics[width=5in]{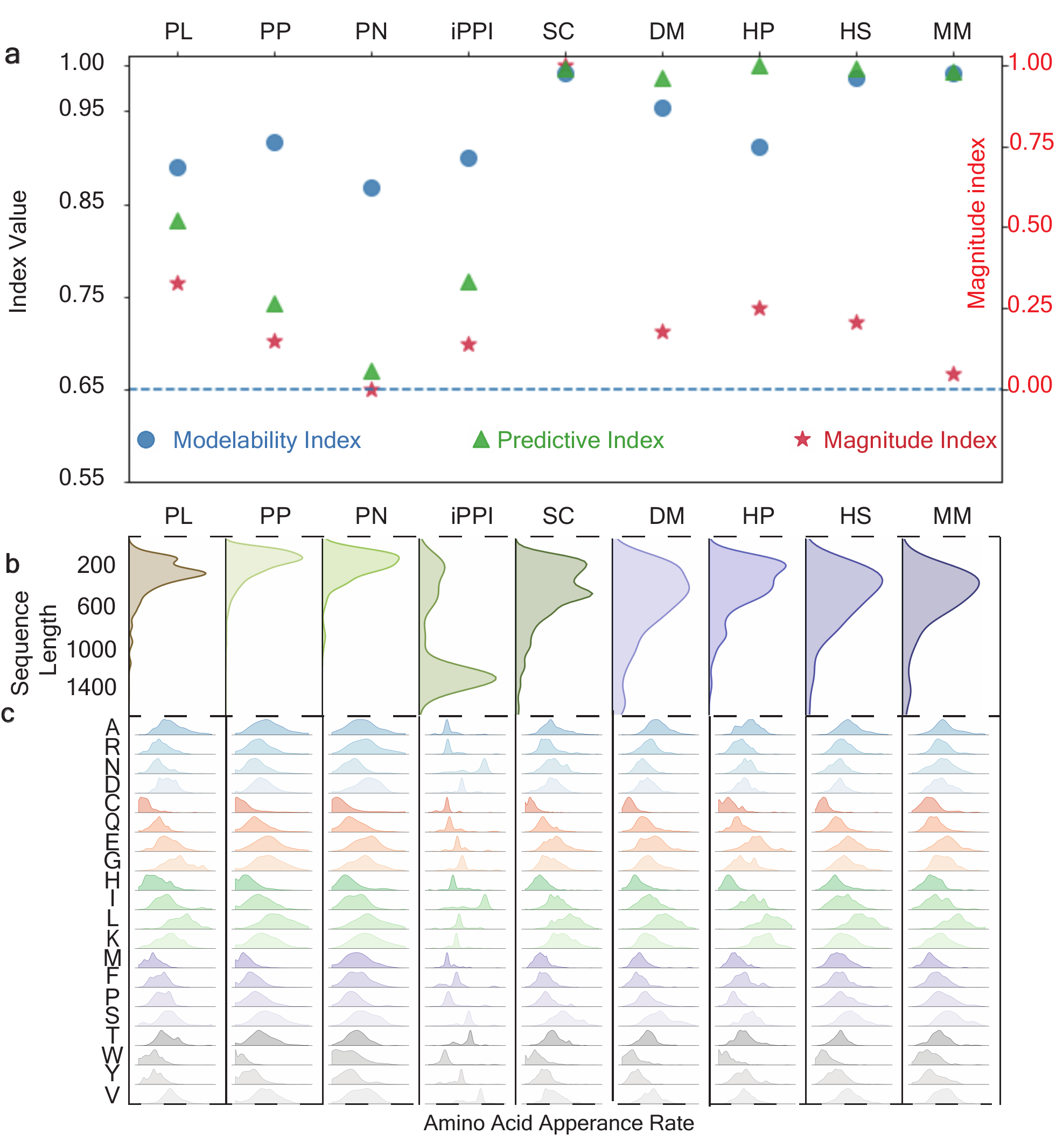} 
	\caption{Analysis of  nine datasets. 
	(a) Modelability index, predictive index, and magnitude index for nine datasets. The left $y$-axis represents  modelability and predictive indices, while the right $y$-axis is the magnitude index. 
	Nine datasets used in our work are four binding affinity regression tasks (i.e., PL, PP, PN, iPPI), and five protein-protein interaction classification tasks, namely SC ( Saccharomyces cerevisiae), DM (Drosophila melanogaster), HP (Helicobacter pylori), HS (Homo sapiens), and MM (Mus musculus).  
	(b) The distribution of sequence length for 9 datasets.     
	(c) The normalized amino acids appearance rate distribution. This subfigure has 9 channels horizontally, corresponding to nine datasets described in (a) and (b). Each channel shows the distribution of 20 types of amino acids appearance rates in sequences of the dataset.}	
	\label{fig:discussion}
\end{figure}

In predicting biomolecular interactions, structure-based approaches are popular and highly accurate when the topological representations of high-quality 3D structures are employed \cite{cang2018representability}. However, their performance depends on the availability of reliable high-resolution experimental structures. Structural docking is a necessary protocol for structure-based approaches when there is no experimental structure available for the interactive complex. Additionally, the power of structure-based methods lies in their ability to accurately capture the geometric information of the interactive complexes. Therefore, the disparity between docked structures and experimental structures will also be inherited by structure-based models. However, no studies have shown that current $K$-embedding models can control this disparity within acceptable tolerances. 
By contrast, our SVS method provides an alternative approach for the study of interactive molecular complexes using only sequence data. It implicitly embeds structural information, flexibility, structural evolution, and diversity in the latent space, which is optimized for downstream models through $K$-embedding  strategies.  It is worth noting that  SVS reaches the same level accuracy as of the best structure-based approach as shown in \autoref{fig:pl}.

Ligand-based virtual screening models also serve as another effective approach that can avoid structure-based docking for evaluating biomolecular interaction with ligands \cite{ripphausen2011state}. However, the current usage of ligand-based models is quite limited as these models in principle can only be applied to target-specific datasets and cannot be used for the screening involving new targets. We showed that by combining target and ligand   deep embeddings via $K$-embedding strategies, SVS gives rise to robust target-unspecific predictions with structure-based accuracy. 

The Biological language processing module and the $K$-embedding  module are two major components in SVS models. Conventionally, the model performance relies on both featurization modules and machine learning algorithms. To solely analyze the quality of the featurization modules, we carry out residue-similarity (R-S) analysis using R-S plot and R-S index \cite{hozumi2022ccp} for classification tasks (\autoref{fig:ppi}b). The R-S analysis describes the quality of features in terms of similarity scores and residue scores as well as the deviation between different classes.

We further analyze SVS behaviors on different datasets in terms of magnitudes and modelability (\autoref{fig:discussion}a) where the basic information of correspondence datasets can be found in Table S3. Three metrics are employed: modelability index,  predictive, and index magnitude index. The modelability index and magnitude index are calculated based on the training data of each dataset, while the predictive index is calculated based on our predictive results on the test data. Note that if our model is tested via cross-validation, then the whole dataset will be calculated for each of the five indices. The predictive index is chosen based on task types: we chose the accuracy score for classification tasks and $R_{\text{p}}$ for regression tasks. The modelability index, which represents the feasibility of our approach on the training data of each dataset, is evaluated by calculating the class-weighted ratio (classification) or the activity cliff (regression) between the nearest-neighbors of samples (\autoref{eq:modelability_cl} and \autoref{eq:modelability_reg}). Previous studies \cite{luque2018study,marcou2016kernel} have suggested that 0.65 is the threshold to separate the modelable and non-modelable datasets. Our model exceeds this threshold in all datasets. In particular, the modelability indices exceed 0.8, which confirms the robustness, stability, and feasibility of our SVS. Our method is compatible with a wide variety of dataset sizes, as shown by the magnitude index, which reflects the corresponding dataset size in proportion to the maximal size of the 9 datasets studied (the maximal data size is 11188). Our analysis shows that there is no significant correlation between the magnitude index and with modelability index or the predictive index, with the only exception being the PN dataset. This dataset, compared to other datasets of the same task (i.e., PL, PP, iPPI datatsets ), has the same level of modelablity index, but with lower levels of the predictive index. We believe that this is because the magnitude index is too small, and this dataset is tested by cross-validation. Therefore, the randomly selected data leads to a void in the feature space, making it difficult for our model to fit this dataset. In conclusion, SVS can be broadly applied for biomolecular predictions and is robust against data size variation.
Moreover, SVS has a strong adaptability to molecules with different sequence compositions. Since proteins were involved in each of our previous numerical experiments, we show the length distribution of protein sequences in each dataset (\autoref{fig:discussion}b) as well as the distribution of amino acids appearance rate in the sequences (\autoref{fig:discussion}c). On average, the sequence lengths of PL, PP, and PN are shorter than those of \textit{Saccharomyces cerevisiae} (SC), \textit{Drosophila melanogaster} (DM), \textit{Helicobacter pylori} (HP), \textit{Homo sapiens} (HS), and \textit{Mus musculus} (MM). This is because samples in the previous datasets are also provided with experimentally determined structures. The availability and reliability of large-size protein structures are subjected to experimental techniques as well as practical considerations, which leads to inevitable systematical bias for structure-based approaches. On the other hand, our SVS models show excellent performances for tasks involving various sequence length distributions. Furthermore, the diversity of the amino acid appearance rate distribution supports the adaptability of our model for tackling different biological tasks, regardless of whether the sequence composition involved has some specificity. In conclusion, our SVS models are robust against sequence length variation and adaptive to biomolecular variability, which reveals the potential of our SVS method as a universal approach for studying biological interactions.

The success of the SVS is due to the use of powerful NLP models, such as LSTM, autoencoder, and particularly Transformers trained with hundreds of millions of molecules. These models extract the constitutional rules of molecules and biomolecules without resorting to molecular property labels.  The proposed SVS will become more powerful as more advanced NLP models become available. 

To showcase the proposed SVS method, we choose nine representative biomolecular interaction datasets involving four regression datasets for protein-ligand binding, protein-protein binding, nucleic acid binding, and ligand inhibition of protein-protein interactions and five classification datasets for the protein-protein interactions in five biological species. SVS  can be applied to the large-scale virtual screening of multiple targets and multiple molecular components without any structural information.

The further application of SVS to mutational virtual screening, directed evolution,  and protein engineering tasks involving biomolecular interactions \cite{qiu2021cluster}  can be done straightforwardly.  Since the protein Transformer was trained with hundreds of millions of protein sequences, this application should offer very competitive results.



\section{Methods}\label{sec_method}

\subsection{Datasets} \label{sec:dataset}
In this study, we used PDBbind-2016 datasets \cite{liu2015pdb} for predicting the protein-ligand binding affinity. The dataset used in protein-protein binding affinity was constructed from PDBbind database \cite{liu2015pdb}. The original PDBbind version 2020 contains binding affinity data of 2852 protein-protein complexes. We selected 1795 samples with only two different sub-chain sequences as shown in Table S5. Furthermore, we also construct the protein-nucleic acid binding affinity dataset from  PDBbind version 2020. However, unlike proteins and ligands, nucleic acids need to be converted to k-mers (in our models, k equals  3) before feeding into the  Transformer model we used. Thus, one unconventional letter (e.g., X, Y) in a sequence will result in k unknown k-mers. In addition, nucleic acids binding to proteins are generally short in length. Therefore, thus unconventional letters in their sequences may completely destroy the context of k-mer representations. For example, a nucleic acid sequence ``ACXTG" will be converted into three 3-mers: ``ACX", ``CXT", and ``XTG". Note that these three 3-mers all contain an ``X", so the  biomolecular language processing model will treat them as unknown tokens, and will not be able to read any useful sequence information. In order to guarantee the effectiveness of sequence information, we apply a stricter excluded criterion: 1) exclude those protein-nucleic acid complexes that their sequence numbers do not equal two; 2)  exclude those protein-nucleic acid complexes that have  unclear labels; 3)  exclude those  protein-nucleic acid complexes that have abnormal letters (normal ones are A, C, T, G) in its nucleic-acid sequences; 4)  exclude those protein-nucleic complexes that whose nucleic acid sequence length is fewer than 6. The resulting  dataset contains 186 protein-nucleic acid complexes as  shown in Table S4.

The original dataset iPPI dataset focuses on ligands thus the availability of PPI targets is obscure and only 31 targets are provided at the family level while 1694 ligands are available. For each protein family, we selected one protein to represent the whole family (e.g., we chose P10415/Q07812 for BCL2/BAK; O60885/P62805 for bromodomain/histone, and O75475/P12497 for ledgf/in.). More specific correspondences can be found in Table S6.

The protein-protein interaction identification involves five benchmark datasets, namely 2434 proteins pairs from \textit{Homo sapiens}, 694 protein pairs from \textit{Mus musculus}, 11188 protein pairs from \textit{Saccharomyces cerevisiae}, 2140 protein pairs from \textit{Drosophila melanogaster}, and 2916 protein pairs from \textit{Helicobacter pylori} \cite{zhao2022protein}. Each dataset consists of an equal quantity of interacting pairs and non-interacting pairs. The interacting protein pairs, serving as positive samples, were collected from the public Database of Interacting Proteins (DIPs) \cite{xenarios2002dip}. Samples with fewer than 50 amino acids and more than 40\% pairwise sequence identity to one another were excluded to reduce fragments and sequence similarity. Negative samples of each dataset were generated by randomly selecting protein pairs in distinct sub-cellular compartments. Proteins from different sub-cellular compartments usually do not interact with each other, and indeed, this construction assures high confidence in identifying negative samples \cite{zhao2022protein}.  

\subsection{$K$-embedding  strategies} \label{sec:k_emb}

For a given  molecular complex   with $m$ molecules, denote  $S_m=\{ s_1,s_2,\cdots,s_m\} (m\ge 2)$ the set of  the corresponding sequences. The set of associated  NLP  1-embeddings is \\ 
$\{\tau_{u_1}^{(1)}({s_1}),\tau_{u_2}^{(1)}({s_2}),\cdots,\tau_{u_m}^{(1)}({s_m}) \}$. Here the subscript ($u_i$) is the embedding dimension, e.g., 512 for the latent space dimension of small molecular Transformer \cite{chen2021extracting}.  
Our goal is to construct an optimal $m$-embedding model    ($\tau_z^{(m)}({S_m})$) from $\{\tau_{u_1}^{(1)}({s_1}),\tau_{u_2}^{(1)}({s_2}),\cdots,\tau_{u_m}^{(1)}({s_m}) \}$, for the complex.   

In general, a $q$-embedding is defined on lower forms as the following formula:
\begin{equation}
	\tau^{(q)}_w({S_q}):= H(\tau^{(r)}_u({S_r}), \tau^{(t)}_v({S_t})), 
\end{equation} 
where $r+t=q$, and $S_r=\{s_{i_1},s_{i_2},\cdots,s_{i_r}\}, S_t=\{s_{j_1},s_{j_2},\cdots,s_{j_t}\}, ~{\rm and }~S_q=\{s_{k_1},s_{k_2},\cdots,s_{k_{q}}\}$ are three subsets of sequences. Here, the $H$ is the integrating function. In this study, we applied \textit{Stack}, \textit{Prod}, and \textit{Diff} based on the homogeneity or heterogeneity of strategies of lower forms as our choices of $H$. 

Specifically, the \textit{Stack} can be defined as follows:
\begin{equation}\label{eq:stack}
	Stack(\tau^{(r)}_u({S_r}),\tau^{(t)}_v({S_t}))=\tau^{(r)}_u({S_r}) \oplus \tau^{(t)}_v({S_t})
\end{equation}
where  $\oplus$ is the direct sum. 

Furthermore, if the lower form strategies are homogenous (i.e., $u=v, s=t$), we can define the \textit{Prod} and \textit{Diff} as follows:

\begin{equation}\label{eq:prod}
	Prod(\tau^{(r)}_u({P_r}),\tau^{(t)}_v({P_t}))= \frac{prod-\mu (prod)}{\sigma (prod)},
\end{equation}
\begin{equation}\label{eq:diff}
	Diff(\tau^{(r)}_u({P_r}),\tau^{(t)}_v({P_t}))= \frac{diff-\mu (diff)}{\sigma (diff)},
\end{equation}
where   $\mu $ and $\sigma$ are the mean value and standard deviation, and 
\begin{equation}
	prod = \tau_u^{(r)}(P_r)\times \tau_v^{(t)}(P_t),
\end{equation}
\begin{equation}
	diff = \tau_u^{(r)}(P_r) - \tau_v^{(t)}(P_t),
\end{equation}
where $\times$ and $-$ is the element-wise product and subtraction, respectively.

In this work, the optimization is made over individual NLP embedding ($\tau_{u_j}^{(1)}(s_j)$), such as Transformer, autoencoder, and LSTM, and all the integrating functions ($H$), i.e., \textit{Stack}, \textit{Prod}, and \textit{Diff}.

\subsection{Machine learning algorithms} \label{sec:models}

We use two set of machine learning algorithms. The first set is the artificial neural networks (ANN), a deep learning algorithm that inspired from the complicated functionality of human brain.  For each task, we use Bayesian optimization \cite{snoek2012practical} to search the best combination of hyperparameters including network size, L2 penalty parameters, learning rate, batch size, and max iteration. The second model is the gradient boost decision tree (GBDT), one of the most popular ensemble methods. GBDT has the advantages of robustness against overfitting, insensitiveness to hyperparameters, effectiveness in the performance, possession of interpretability. GBDT was mainly used to implement regression tasks. The hyperparameters including ``n\_estimators, max\_depth, min\_sample\_split, subsample, max\_features" are chosen based on the data size and embedding dimensions of each task. The optimization strategies used in our study are presented in Table S2.

\subsection{Bayesian optimization for ANN hyperparameter  tuning} \label{sec:bayes_opt}

  Bayesian optimization is a popular approach to sequentially optimize hyperparameters of machine learning algorithms. The Bayesian optimization is to maximize a black-box function $f(x)$ in a space $\mathcal{S}$:
\begin{equation}
	x^{*}=\arg\max_{x\in \mathcal{S}}f(x),
\end{equation}
In the hyperparameter optimization, $\mathcal{S}$ can be regarded as the search space of hyperparameters, $x^{*}$ is the set of optimal hyperparameters, and $f(x)$ is an evaluating metric for machine learning performance. 

Given $t$ data points $X_t=(x_1,x_2,\cdots,x_t)$ and their values of evaluating matrics $Y_t=(y_1,y_2,\cdots,y_t)$, Gaussian process can model the landscape of $f$ on the entire space $\mathcal{S}$ by fitting $(X_t,Y_t)$ \cite{williams2006gaussian}. At any novel point $x$, $f(x)$ is modeled by a Gaussian posterior distribution: $p(f(x)|X_t,Y_t)\sim\mathcal{N}(\mu_t(x),\sigma^2_t(x))$, where $\mu_t(x)$ is mean and $\sigma$ is the standard deviation of $f(x)$ predicted by Gaussian process regression:
\begin{equation}
	\begin{aligned}
		&\mu_t(x)=K(x,X_t) \left[K(X_t,X_t)+\epsilon_n^2I \right]^{-1}Y,\\
		&\sigma^2_t(x)=k(x,x)-K(x,X_t)\left[K(X_t,X_t)+\epsilon_n^2I\right]^{-1}K(x,X_t)^T.
	\end{aligned}
\end{equation}
Here $k$ is the kernel function, $K(x,X_t)$ is a row vector of kernel evaluations between $x$ and the elements of $X_t$ with$[K(x,X_t)]_i=k(x,x_i)$, and $K(X_t,X_t)$ is the kernel matrix with $[K(X_t,X_t)]_{ij}=k(x_i,x_j)$. $\epsilon_n$ is the noise term, which is learned from the regression. 

In Bayesian optimization, both predicted mean and standard deviation are used for the decision making for the next evaluating data point. One can either pick the point maximize the mean values of $f(x)$ for a greedy search, or pick the point with the largest standard deviation to gain new knowledge and improve the Gaussian process accuracy on $f(x)$ landscape. The greedy search may largely maximize $f(x)$ in a few iterations and the exploration of uncertain points can benefit for   long-term iterations. To balance such a exploitation-exploration trade-off, an acquisition function, $\alpha(x)$, needs to be picked. The decision for the next evaluating point $x_n$ is picked such that it maximizes the acquisition function
\begin{equation}
	x_n=\arg\max_{x\in\mathcal{S}}\alpha(x).
\end{equation}
 In this study, we used the upper confidence bound (UCB) acquisition which can handle the trade-off and it has a fast convergent rate \cite{srinivas2009gaussian} for the black-box optimization.

\subsection{Evaluation metrics}

In addition to the evaluation metrics given in the Supplementary Information (Equation S1 to Equation S7), 
R-S scores, R-S index, and modelability index are described below. 


\subsubsection{R-S scores}

Residue-Similarity (R-S) plot is a new kind of visualization and analysis method that can be applied to an arbitrary number of classes proposed by Hozumi et al. \cite{hozumi2022ccp}. An R-S plot evaluates each sample of given data by two components, the residue and similarity scores. For given dataset $\{(x_m,y_m)|x_m \in R^{N},y_m\in Z_L\}_{m=1}^{M}$, the residue score and the similarity score of a sample $(x_m,y_m)$ are defined as follows:
	\begin{equation} \label{eq:r-score}
		R_m := R(x_m) =\frac{\sum_{x_j \notin C_l}||x_m-x_j||}{\max\limits_{x_m\in C_l}(\sum_{x_j\notin C_l}||x_m-x_j||)},
	\end{equation}
\begin{equation} \label{eq:s-score}
			S_m := S(x_m) = \frac{1}{|C_l|}\sum_{x_j\in{C_l}}(1-\frac{||x_m-x_j||}{d_{\max}}),
\end{equation}
where $l=y_m$, $C_l=\{x_m|y_m=l\}$, and $d_{\max}=\max\limits_{x_i,x_j\in C_l}||x_i-x_j||$. Note that $0\leq R_m\leq1$ and $0\leq S_m \leq 1$. If a sample is far from other classes, it will have a larger residue score; if a sample is well-clustered, it will have a larger similarity score.

 The Class residue index (CRI) and class similarity index (CSI) for the $l$-th class can be defined as $\text{CRI}_l=\frac{1}{|C_l|}\sum_m R_m$ and $\text{CSI}_l = \frac{1}{|C_l|}\sum_m S_m$. Then the class-independent residue index (RI) and similarity index (SI) can be defined:
\begin{equation}
\text{RI}:=\frac1L\sum_l\text{CRI}_l,
\end{equation}
\begin{equation} 
\text{SI}:=\frac1L\sum_l\text{CSI}_l.
\end{equation}
Then the R-S indices which can give a class-independent evaluation of the deviation R- and S- scores \cite{hozumi2022ccp} can be defined:
\begin{equation} \label{eq:rsi}
	\text{RSI}:= 1-|\text{RI}-\text{SI}|
\end{equation}
Note that RSI range from 0 to 1 and a low RSI indicates a large deviation between the R-score and S-score.

\subsubsection{Modelability}

The modelability index is defined independently for classification tasks and regression tasks, namely $\text{MODI}_\text{cl}$ and $\text{MODI}_\text{reg}$, respectively, defined as follows \cite{luque2018study,marcou2016kernel}:
\begin{equation} \label{eq:modelability_cl}
	\text{MODI}_\text{cl} = \frac{1}{L}\sum_{i=1}^{L}\frac{N_i}{M_i},
\end{equation}
\begin{equation} \label{eq:modelability_reg}
	\text{MODI}_\text{reg} = 1-\frac{1}{M}\sum_{i=1}^{M}\frac{1}{K_i}\sum_{j \in C^1_i}|y_i-y_j|,
\end{equation}
where $L$ represents the number of classes, $N_i$ is the count of samples in the $i$-th class whose first nearest neighbor is also in the $i$-th class, $M_i$ is the number of samples in the $i$-th class, $M$ is the total number of samples, $C^1_i$ is the 1-nearest neighbor of $i$-th sample, $K_i$ is the count of samples in $C^1_i$ except the $i$-th sample, and $y_i$ represents the $i$-th samples' normalized label.

\subsection*{Data Availability}
 All datasets are available at \hyperlink{https://weilab.math.msu.edu/DataLibrary/2D/}{https://weilab.math.msu.edu/DataLibrary/2D/}.

\subsection*{Code Availability}
 The source codes are available at \hyperlink{https://github.com/WeilabMSU/MTN}{https://github.com/WeilabMSU/SVS}.

\section*{Acknowledgments}
This work was supported in part by NIH grants  R01GM126189 and  R01AI164266, NSF grants DMS-2052983,  DMS-1761320, and IIS-1900473,  NASA grant 80NSSC21M0023,  MSU Foundation,  Bristol-Myers Squibb 65109, and Pfizer.

\end{document}


\title{ Supplementary Information for  \\                                         
 SVSBI: Sequence-based virtual screening of biomolecular
interactions}
\author{Li Shen$^1$,  Hongsong Feng$^{1 }$  ,  Yuchi Qiu$^{1 }$ 
	and Guo-Wei Wei$^{1,2,3}$\footnote{
		Corresponding author.		E-mail: weig@msu.edu} \\
	$^1$ Department of Mathematics, \\
	Michigan State University, MI 48824, USA.\\
	$^2$ Department of Electrical and Computer Engineering,\\
	Michigan State University, MI 48824, USA. \\
	$^3$ Department of Biochemistry and Molecular Biology,\\
	Michigan State University, MI 48824, USA. \\
}

    \maketitle
		\pagenumbering{roman}
{\setcounter{tocdepth}{4} \tableofcontents}

\newpage
\pagenumbering{arabic}

\section{Additional evaluation metrics}
		In this study, we used, accuracy (ACC), precision (Pre), sensitivity (Se), Matthews correlation coefficient (MCC), and F1-score, and geometry mean score (GM) to evaluate the performance of our models for classification problems. These evaluation metrics are defined as below:

		\begin{equation} \label{eq:acc}
			ACC = \frac{TP+TN}{TP+TN+FP+FN}
		\end{equation}
		\begin{equation} \label{eq:pre}
			Pre = \frac{TP}{TP+FP}
		\end{equation}
		\begin{equation} \label{eq:se}
			Se = \frac{TP}{TP+FN}
		\end{equation}
		\begin{equation}\label{eq:mcc}
			MCC = \frac{TP\times TN -FP\times FN}{\sqrt{(TP+FP)(TP+FN)(TN+FP)(TN+FN)}}
		\end{equation}
		\begin{equation} \label{eq:f1}
			\text{F}1\text{-score} =\frac{2\times Pre \times Se}{Pre+Se},
		\end{equation}
		where  TP (True positive) indicates the number of positive samples that are predicted as positive; false negative (FN$_i$) indicates the number of negative samples that are predicted as negative; true negative (TN$_i$) indicates the number of negative samples that are predicted as negative; and false positive (FP$_i$) indicates the number of negative samples that are predicted as negative.
		
		We also used receiver operating characteristic (ROC) curve \cite{fawcett2006introduction} and Area under ROC curve (AUC) to evaluate of our model at all decision thresholds. An ROC curve plots True-positive Rate (TPR) vs. False-positive Rate at different decision thresholds. Lowering the decision threshold will classifies more instances as positive, thus increasing both TPR and FPR. AUC measures the entire two-dimensional area underneath the ROC curve, ranging from 0 to 1. The higher the AUC, the better the performance of the model at distinguishing the positive and negative classes.
	
		Furthermore, the Pearson correlation coefficient ($R_\text{p}$) is used to evaluate our models in regression tasks, and it is defined as follows:
		\begin{equation} \label{eq:rp}
			R_\text{p} = \frac{\sum(x_i-\bar{x})(y_i-\bar{y})}{\sqrt{\sum(x_i-\bar{x})^2(y_i-\bar{y})^2}},
		\end{equation}
where $x_i$ is the value of the $x$ variable in the $i$-th sample, $y_i$ is the value of $y$ variable in the $i$-th sample, $\bar{x}$ is mean value of the $x$ variable and $\bar{y}$ is mean value of the $y$ variable. Also, the root mean squared error (RMSE) is applied,and it is defined as below,
       \begin{equation} \label{eq:rmse}
            	\text{RMSE} = \sqrt{\frac1n\sum_{i=1}^{n}(y_i-\hat{y_i})^2},
       \end{equation}
where $y_i$ and $\hat{y_i}$ are predicted and true value of the $i$-th sample, respectively.

\section{Additional results}

\begin{table}[H] 
	\centering
	
	\begin{tabular*}{\textwidth}{c@{\extracolsep{\fill}}ccccccc}
		\hline
		\makecell[c]{\textbf{Dataset}}& \textbf{Method}&\textbf{Accuracy}&\textbf{Precision}&\textbf{Sensitivity}&\textbf{F1-score}&\textbf{MCC}\\
		\hline		
		\makecell[c]{\multirow{3}{*}{\textit{H. sapiens}}}&SVM-NVDT\cite{zhao2022protein}&0.9541&0.9159&\textbf{1.0000}&0.9561&0.9121\\
		~&RF-NVDT\cite{zhao2022protein}&0.9541&0.9159&\textbf{1.0000}&0.9541&0.9121\\
		~&\textbf{SVS}&\textbf{0.9959}&\textbf{1.0000}&0.9918&\textbf{0.9959}&\textbf{0.9918}\\
		\hline
		\makecell[c]{\multirow{3}{*}{\textit{M. musculus}}}&SVM-NVDT\cite{zhao2022protein}&0.9643&0.9310&0.8971&0.9474&0.8971\\
		~&RF-NVDT\cite{zhao2022protein}&0.9483&\textbf{1.0000}&0.8966&0.9014&0.9455\\
		~&\textbf{SVS}&\textbf{0.9928}&0.9942&\textbf{0.9914}&\textbf{0.9928}&\textbf{0.9856}\\
		\hline
		\makecell[c]{\multirow{14}{*}{\textit{S. cerevisiae}}}&SVM-AC\cite{guo2008using}&0.8735&0.8782&0.8730&0.8734&0.7509\\
		&kNN-CTD\cite{yang2010prediction}&0.8615&0.9024&0.8103&0.8539&NA\\
		&EELM-PCA\cite{you2013prediction}&0.8699&0.8759&0.8615&0.8686&0.7736\\
		&SVM-MCD\cite{you2014prediction}&0.9136&0.9194&0.9067&0.9130&0.8421\\
		&RF-PR-LPQ\cite{wong2015detection}&0.9392&0.9645&0.9110&0.9380&0.8856\\
		&MLP\cite{du2017deepppi}&0.9443&0.9665&0.9206&0.9430&0.8897\\
		&PCVMZM\cite{wang2017pcvmzm}&0.9448&0.9392&0.9513&NA&0.8959\\
		&DPPI\cite{hashemifar2018predicting}&0.9455&0.9668&0.9224&0.9441&NA\\
		&LightGBM\cite{chen2019lightgbm}&0.9507&0.9782&0.9221&NA&0.9030\\
		&PIPR\cite{chen2019multifaceted}&0.9709&0.9700&0.9717&0.9709&0.9417\\
		&StackPPI\cite{chen2020improving}&0.9464&0.9633&0.9281&NA&0.8934\\
		&TAGPPI\cite{song2022learning}&0.9781&0.9810&0.9826&0.9780&0.9563\\
		&SVM-NVDT\cite{zhao2022protein}&0.9920&0.9935&0.9903&NA&0.9839\\
		&\textbf{SVS}&\textbf{0.9964}&\textbf{0.9998}&\textbf{0.9998}&\textbf{0.9964}&\textbf{0.9929}\\
		\hline
		\makecell[c]{\multirow{12}{*}{\textit{H. pylori}}}&HKNN\cite{nanni2005hyperplanes}&0.8400&0.8400&0.8600&NA&NA\\
		&DCT+SMR\cite{huang2015using}&0.8674&0.8701&0.8643&NA&0.7699\\	
		&S-products\cite{martin2005predicting}&0.8340&0.8570&0.7990&NA&NA\\
		&P-bootstrap\cite{bock2003whole}&0.7580&0.6980&0.8020&NA&NA\\
		&SVM-MCD\cite{you2014prediction}&0.8491&0.8612&0.8324&NA&0.7440\\
		&LightGBM\cite{chen2019lightgbm}&0.8779&0.8787&0.8772&NA&0.7561\\
		&XGBoost\cite{chen2020improving}&0.8927&0.9037&0.8793&NA&0.7859\\
		&PCVMZM\cite{wang2017pcvmzm}&0.9125&0.9006&0.9205&NA&0.8404\\
		&DeepPPI\cite{du2017deepppi}&0.8623&0.8432&0.8944&NA&0.7263\\
		&StackPPI\cite{chen2020improving}&0.8927&0.9037&0.8793&NA&0.7859\\
		&SVM-NVDT\cite{zhao2022protein}&0.9856&0.9836&0.9877&NA&0.9712\\
		&\textbf{SVS}&\textbf{0.9993}&\textbf{1.0000}&\textbf{0.9986}&\textbf{1.0000}&\textbf{0.9986}\\
		\hline
		\makecell[c]{\multirow{4}{*}{\textit{D. melanogaster}}}&NVD\cite{zhao2017establishing}&0.7978&0.8487&0.7247&0.7818&0.6020\\
		&RFDC\cite{najafabadi2008sequence}&0.8904&0.8971&0.8820&0.8895&0.7810\\
		&SVM-NVDT\cite{zhao2022protein}&0.9494&0.9762&0.9213&0.9480&0.9003\\
		&\textbf{SVS}&\textbf{0.9869}&\textbf{1.0000}&\textbf{0.9738}&\textbf{0.9867}&\textbf{0.9742}\\
		\hline
	\end{tabular*}
\begin{flushleft}
\scriptsize{Note: NA means not available.}
\end{flushleft}
\caption{Comparison of our method with previous models for protein-protein interaction identification. The best scores of each  dataset are marked in bold.}\label{table:ppid}
\end{table}
\begin{figure}[H] 
	\centering
    \includegraphics[width=\textwidth]{Figures\s_auc_ppid}

    \caption{The ROC-AUC curve of MTN models for protein-protein interaction identification. Note that all $y$-axes start from 0.990.}
    \label{fig:auc_curve_ppid}
\end{figure}

\section{Model Optimization Parameters}
\begin{table}[H] 
	\begin{tabular*}{\textwidth}{c@{\extracolsep{\fill}}ccc}
		\hline
		Model&Parameters&choices&Remark\\
		\hline
		\multirow{6}{*}{GBDT}&n\_estimators&10000&\\
		~&learning\_rate&0.01&\\
		~&max\_feature&`sqrt'&\\
		~&max\_depth&i=7, j=8, k=9&i if train size $<$ 1000\\
		~&min\_samples\_split&i=3, j=4, k=7&j if 1000$\leq$train size$<$5000\\
		~&sbusample&i=0.7, j=0.5, k=0.3&k if train size $\geq$5000\\
		\hline
		\multirow{6}{*}{MLP}&N\_layer&1,2,3,4,5,6&~\\
		~&N\_neuron&16, 32, 64, 128, 512, 1024&~\\
		~&batch\_size&32, 64, 128, 256&8, 16, 32 for small size datasets\\
		~&max\_iter&20, 40, 80, 160, 320&~\\
		~&alpha&0.01, 0.001, 0.0001, 0.00001&L2 regularization term\\
		~&learning\_rate&0.01, 0.001, 0.0001&~\\
		\hline
	\end{tabular*}
\caption{GBDT and ML parameters used in our study. All parameters not covered in the table use default values. We used Bayesian optimization to search the best combination of parameters for MLP models. }
\label{table:model_opt}
\end{table}

\section{Datasets} \label{sec_sup_fig}

In this section, we   display overview of basic information and benchmark procedure for datasets used in our study in Table \ref{table:dataset}. Then, we present our self-constructed datasets used for protein-protein and protein-nucleic acid binding affinity prediction. Finally, we   show our selections of representations of PPI family for iPPI dataset.
\begin{table}[H]
	\begin{tabular}{cccccc}
		\hline
	S/N&name&source&train size&test size&Repetition\\\hline
	1&PL&PDBbind2016 \cite{berman2000protein}&3772&285&10\\\hline
	2&PP&PDBbind2020 \cite{berman2000protein}&1795&10-fold cross-validation&20\\\hline
	3&PN&PDBbind2020 \cite{berman2000protein}&186&10-fold cross-validation&20\\\hline
	4&iPPI&Rodrigues et al. \cite{rodrigues2021pdcsm}&1694&10-fold cross-validation&50\\\hline
	5&\textit{S. cerevisiae}&Zhao et al. \cite{zhao2022protein}&11188&5-fold cross-validation&1\\\hline
	6&\textit{D. melanogaster}&Zhao et al. \cite{zhao2022protein}&2140&5-fold cross-validation&1\\\hline
	7&\textit{H. pylori}&Zhao et al. \cite{zhao2022protein}&2916&5-fold cross-validation&1\\\hline
	8&\textit{H. sapiens}&Zhao et al. \cite{zhao2022protein}&2434&5-fold cross-validation&1\\\hline
	9&\textit{M. musculus}&Zhao et al. \cite{zhao2022protein}&694&5-fold cross-validation&1\\\hline
	\end{tabular}
\caption{Overview of basic information of all datasets used in this study. The number of repetitions follows the benchmark procedure of previous methods (if present).}
\label{table:dataset}
\end{table}

Furthermore, we present the PP (protein-protein interaction) and PN (protein-nucleic acid interaction) datasets that we constructed in the task of binding affinity predictions. The original PDBbind2020 database provides the labels in term of dissociation constant ($\text{K}_d$), inhibitor constant ($\text{K}_i$), and half maximal inhibitory concentration ($\text{IC}_{50}$), we transform them into Gibbs free energy with the following equations:
\begin{equation} \label{eq:free_energy}
	\Delta G = -RTpK,
\end{equation}
where $pK$ is $log_{10}K$ where $K$ represents $\text{K}_d$ or $\text{K}_i$. And $\text{IC}_{50}$ can be approximately converted to $K_i$ by the equation $K_i=\text{IC}_{50}/2$. $R$ and $T$ are the gas constant and temperature, respectively. At the room temperature, this formula becomes $\Delta G = -1.3633pK$.

\newpage
\begin{longtable}[H]{|c|c|c||c|c|c|} 
	\hline
	PDB entry&label&$\Delta G$&PDB entry&label&$\Delta G$	\\\hline
	\endhead
	1HVO&Kd=11uM&-6.76&
	1HVN&Kd=5uM&-7.227\\\hline
	1WET&Kd=1.5uM&-7.94&
	1BDH&Kd=900nM&-8.242\\\hline
	2PUF&Kd=2.0uM&-7.769&
	1QP4&Kd=10nM&-10.906\\\hline
	1QPZ&Kd=2.6nM&-11.704&
	1QQB&Kd=2000nM&-7.769\\\hline
	1QP0&Kd=35nM&-10.165&
	1QQA&Kd=400nM&-8.722\\\hline
	1QP7&Kd=780nM&-8.327&
	1DH3&Kd=1.4nM&-12.07\\\hline
	1FJX&Kd=1.4nM&-12.07&
	1JJ4&Kd=1.6nM&-11.991\\\hline
	1J75&Kd=35nM&-10.165&
	1FYM&Kd=40nM&-10.086\\
\end{longtable}
\begin{longtable}[H]{|c|c|c||c|c|c|} 
	\hline
	PDB entry&label&$\Delta G$&PDB entry&label&$\Delta G$	\\\hline
	\endhead
	2NYZ&Kd=0.9nM&-12.332&3OAK&Kd=169nM&-9.232\\\hline
	2ZXX&Kd=4.4nM&-11.392&4U2X&Kd=1nM&-12.27\\\hline
	3BRV&IC50=47nM&-9.99&2FDB&Kd=155nM&-9.284\\\hline
	4Z80&Kd=12nM&-10.798&3M63&Kd=175nM&-9.212\\\hline
	4A49&Kd=42uM&-5.967&3ETB&Kd=29pM&-14.366\\\hline
	2Q5D&Kd=15.6nM&-10.643&1E96&Kd=2.7uM&-7.592\\\hline
	5B75&Kd=10uM&-6.816&5YIP&Kd=3.7nM&-11.495\\\hline
	1BUH&Kd=77nM&-9.698&3C4M&Kd=0.98uM&-8.192\\
\end{longtable}

\begin{table}[H]
	\centering
	\begin{tabular}{|c|c|}
		\hline
		PPI family(s)& Uniprot ID\\\hline
		ttr&P02766\\\hline
		hif1a/p300&Q16665/Q09472\\\hline
		wdr5/mll&P61964/Q03164\\\hline
		RUNX1/CBFb&Q01196/Q13951\\\hline
		cd80/cd28&P33681/P10747\\\hline
		ras/sos1&P01112/Q07889\\\hline
		BRI1&Q942F3\\\hline
		VEGF/VEGFR&P15692/P17948\\\hline
		il2/il2r&P60568/P01589\\\hline
		CD40L&trimer : P29965\\\hline
		Tubulin&P02554\\\hline
		PCNA trimer&P12004\\\hline
		cyclophilins&P62937\\\hline
		bromodomain/histone&O60885/P62805\\\hline
		integrins&P56199\\\hline
		SPIN1/Q5TEC6&Q9Y657/Q5TEC6\\\hline
		cd4/gp120&P01730/P04578\\\hline
		bcl2-like/bak&P10415/Q07812\\\hline
		E2/E1&P06790/P06789\\\hline
		MLLT1/H3&Q03111/P68431\\\hline
		annexin/s100&P07355/P60903\\\hline
		SOD1&P00441\\\hline
		brd2/ack&P25440/Q07912\\\hline
		TNF&P01375\\\hline
		WDR5/MLL&P61964/Q03164\\\hline
		xiap/smac&P98170/Q9NR28\\\hline
		menin/mll&O00255/Q03164\\\hline
		Beta-catenin/TCF-4&P35222/P15884\\\hline
		CRM1/Rev&O14980/Q9UBZ9\\\hline
		mdm2-like/p53&Q00987/P04637\\\hline
		NRP/VEGF&Q9QWJ9/P15692\\\hline
		fkbp1a/fk506&P62942/Q02790\\\hline
		lfa/icam&P20701/P05362\\\hline
		stat3&P40763\\\hline
		ledgf/in&O75475/P12497\\\hline
		UPAR/UPA&Q03405/P00749\\\hline
		Myc/Max&P01106/P61244\\\hline
		WD40/H3&O75530/P68431\\\hline
		CaM/CaMBD2&P07463/Q9H2S1\\\hline
		53BP1/H4&Q12888/Q12888\\\hline
		Rac1/GEFs&P63000/Q8SSW7\\\hline
	\end{tabular}
	\caption{Representations of protein families. The selected proteins are provided with Uniprot entries.}
	\label{table:ppid_represent}
\end{table}



	\bibliographystyle{abbrv}
    \bibliography{Reference/ref}